# The Gravitational Field Equations of a Twisted Skyrmion String: Numerical Solution

**Miftachul Hadi**[1, 2, 4, 5, a], **Malcolm Anderson**[1, b], **Andri Husein**[3]

[1] Department of Mathematics, Universiti Brunei Darussalam,
   NEGARA BRUNEI DARUSSALAM

[2] Physics Research Centre, Indonesian Institute of Sciences (LIPI), INDONESIA

[3] Alumnus of Physics Department of Post Graduate Program
   Sebelas Maret University, INDONESIA

[4] School of Physics, Ulsan National Institute of Science and Technology,
   SOUTH KOREA

[5] Institute of Modern Physics, Chinese Academy of Sciences, Lanzhou, CHINA

E-mail: itpm.id@gmail.com

**Abstract.** We study nonlinear sigma model, especially Skyrme model with twist: twisted Skyrmion string where twist term, $mkz$, is indicated in vortex solution. To add gravity, we replace $g^{\mu\nu}$ in Lagrangian system with a space-time metric tensor, $g^{\mu\nu}$ which in view of the time-independence and cylindrical symmetry of the assumed vortex solution is taken to be a function of $r$ alone. We use ode45 for numerical calculation, i.e. a tool box in Matlab to solve coupled Einstein field equations which have ordinary differential equations (ODE) form.

## 1. Introduction to Nonlinear Sigma Model

The Lagrangian density of a free (without potential) nonlinear sigma model on a Minkowski background space-time is defined to be [1]

$$L = \frac{1}{2\lambda^2} \gamma_{AB}(\phi) \eta^{\mu\nu} \partial_\mu \phi^A \cdot \partial_\nu \phi^B \tag{1}$$

where $\gamma_{AB}(\phi)$ is the field metric, $\eta^{\mu\nu}$ is the Minkowski metric tensor, $\lambda$ is a scaling constant with dimensions of (length/energy)$^{1/2}$ and $\phi = \phi^A$ is the collection of fields.

One of the most interesting examples of a $O(N)$ nonlinear sigma model, due to its topological properties, is the $O(3)$ nonlinear sigma model in (1+1)D, with the Lagrangian density

$$L = \frac{1}{2\lambda^2} \eta^{\mu\nu} \partial_\mu \phi \cdot \partial_\nu \phi \tag{2}$$

---

[1,a] Miftachul Hadi (itpm.id@gmail.com)
[1,b] Malcolm Anderson (malcolm.anderson@ubd.edu.bn)
[3] Andri Husein (andrihusein08@gmail.com)

where $\mu$ and $\nu$ range over $\{0,1\}$, and $\phi = (\phi^1, \phi^2, \phi^3)$, subject to the constraint $\phi \cdot \phi = 1$, where the dot ($\cdot$) denotes the dot product on real coordinate space of three dimensions, $R^3$.

A simple representation of field, $\phi$, (in the general time-dependent case) is [2]

$$\phi = \begin{pmatrix} \sin f(t,\vec{r}) \sin g(t,\vec{r}) \\ \sin f(t,\vec{r}) \cos g(t,\vec{r}) \\ \cos f(t,\vec{r}) \end{pmatrix} \tag{3}$$

where $f$ and $g$ are scalar functions on the background space-time, with Minkowski coordinates $x^\mu = (t,\vec{r})$.

If a Skyrme term is added to $O(3)$ nonlinear sigma model, the result is a modified Lagrangian density, i.e. Skyrme model [2]

$$L = \frac{1}{2\lambda^2}\eta^{\mu\nu}\partial_\mu\phi \cdot \partial_\nu\phi - K_s\eta^{\kappa\lambda}\eta^{\mu\nu}(\partial_\kappa\phi \times \partial_\mu\phi) \cdot (\partial_\lambda\phi \times \partial_\nu\phi) \tag{4}$$

where the Skyrme term is the second term on the right hand side of [4].

## 2. The Gravitational Field of a Twisted Skyrmion String

We are interested in constructing space-time generated by a twisted Skyrmion string. Without gravity, the Lagrangian density of the system is $L_1$, as given in equation below [3]

$$L_1 = \frac{1}{2\lambda^2}\left(\eta^{\mu\nu}\partial_\mu f\partial_\nu f + \sin^2 f \eta^{\mu\nu}\partial_\mu g\partial_\nu g\right) \\ - K_s\left[2\sin^2 f\left(\eta^{\mu\nu}\partial_\mu f\partial_\nu f\right)\left(\eta^{\kappa\lambda}\partial_\kappa g\partial_\lambda g\right) - 2\sin^2 f\left(\eta^{\mu\nu}\partial_\mu f\partial_\nu g\right)^2\right] \tag{5}$$

To add gravity, we replace $\eta^{\mu\nu}$ in $L_1$ (5) with a space-time metric tensor, $g^{\mu\nu}$, which in view of the time-independence and cylindrical symmetry of the assumed vortex solution is taken to be a function of $r$ alone.

Metric tensor, $g^{\mu\nu}$, is of course the inverse of the covariant metric tensor, $g_{\mu\nu}$, of the space-time where $g^{\mu\nu} = (g_{\mu\nu})^{-1}$. We use a cylindrical coordinate system $(t, r, \theta, z)$, where $t$ and $z$ have unbounded range, $r \in [0, \infty)$ and $\theta \in [0, 2\pi)$. The metric tensor with its components can be written as a matrix below

$$g_{\mu\nu} = \begin{pmatrix} g_{tt} & 0 & 0 & 0 \\ 0 & g_{rr} & 0 & 0 \\ 0 & 0 & g_{\theta\theta} & g_{\theta z} \\ 0 & 0 & g_{z\theta} & g_{zz} \end{pmatrix} \tag{6}$$

where the components of metric tensor are all functions of $r$, and the presence of the off-diagonal components $g_{\theta z} = g_{z\theta}$ reflects the twist in the space-time.

The Lagrangian we will be using is [3]

$$L_2 = \frac{1}{2\lambda^2}\left(g^{\mu\nu}\partial_\mu f\partial_\nu f + \sin^2 f g^{\mu\nu}\partial_\mu g\partial_\nu g\right) \\ - K_s\left[2\sin^2 f\left(g^{\mu\nu}\partial_\mu f\partial_\nu f\right)\left(g^{\kappa\lambda}\partial_\kappa g\partial_\lambda g\right) - 2\sin^2 f\left(g^{\mu\nu}\partial_\mu f\partial_\nu g\right)^2\right] \tag{7}$$

where

$$f = f(r), \qquad g = n\theta + mkz \tag{8}$$

We need to solve:
  i. the Einstein field equations

$$G_{\mu\nu} = -\frac{8\pi G}{c^4} T_{\mu\nu} \tag{9}$$

where the stress-energy tensor of the vortex, $T_{\mu\nu}$, is defined by

$$T_{\mu\nu} \equiv 2\frac{\partial L_2}{\partial g^{\mu\nu}} - g_{\mu\nu} L_2 \tag{10}$$

and

$$G_{\mu\nu} = R_{\mu\nu} - \frac{1}{2} g_{\mu\nu} R \tag{11}$$

with $R_{\mu\nu}$ the Ricci tensor and

$$R = g_{\mu\nu} R^{\mu\nu} = g^{\mu\nu} R_{\mu\nu} \tag{12}$$

the Ricci scalar;.

ii. the field equations for $f$ and $g$

$$\nabla^\mu \frac{\partial L_2}{\partial(\partial f/\partial x^\mu)} = \frac{\partial L_2}{\partial f}, \qquad \nabla^\mu \frac{\partial L_2}{\partial(\partial g/\partial x^\mu)} = \frac{\partial L_2}{\partial g} \tag{13}$$

However, the field equations for $f$ and $g$ are in fact redundant, as they are satisfied identically whenever the Einstein field equations are satisfied, by virtue of the Bianchi identities (i.e. permuting of the covariant derivative of the Riemann tensor) $\nabla_\mu G^\mu_\nu = 0$. So, only the Einstein field equations will be considered [3].

To simplify the Einstein field equations, we first choose a gauge condition that narrows down the form of the metric tensor. The gauge condition preferred here is that [3]

$$g_{\theta\theta} g_{zz} - (g_{\theta z})^2 = r^2 \tag{14}$$

The geometric significance of this choice is that the determinant of the 2-metric tensor projected onto the surfaces of constant $t$ and $z$ is $r^2$, and so the area element on these surfaces is just $rdrd\theta$.

As a further simplification, we write $g_{tt} = A^2$, $g_{rr} = -B^2$, $g_{\theta\theta} = -C^2$, $g_{\theta z} = \omega$, where $A$, $B$, $C$, $\omega$ are functions of $r$ and so $g_{zz} = -\left(\frac{r^2 + \omega^2}{C^2}\right)$. The metric tensor, $g_{\mu\nu}$, therefore has the form

$$g_{\mu\nu} = \begin{pmatrix} A^2 & 0 & 0 & 0 \\ 0 & -B^2 & 0 & 0 \\ 0 & 0 & -C^2 & \omega \\ 0 & 0 & \omega & -\left(\frac{r^2+\omega^2}{C^2}\right) \end{pmatrix} \tag{15}$$

Substitute (6), (8) into (7), we obtain $L_2$ as below [4]

$$L_2 = \frac{1}{2g_{rr}} \left[ \frac{1}{\lambda^2} + \frac{4K_s \sin^2 f}{r^2} \left( n^2 g_{zz} - 2nmk g_{\theta z} + m^2 k^2 g_{\theta\theta} \right) \right] \left(\frac{\partial f}{\partial r}\right)^2 \\ + \frac{\sin^2 f}{2\lambda^2 r^2} \left( n^2 g_{zz} - 2nmk g_{\theta z} + m^2 k^2 g_{\theta\theta} \right) \tag{16}$$

From (10), (15) and (16), we derive the non-zero components of the stress-energy tensor, $T_{\mu\nu}$, as below

$$T_{tt} = \frac{A^2}{2B^2}\left(\frac{1}{\lambda^2} + \frac{4NK_s \sin^2 f}{r^2}\right)\left(\frac{\partial f}{\partial r}\right)^2 - \frac{A^2 N \sin^2 f}{2\lambda^2 r^2} \tag{17}$$

$$T_{rr} = \left(\frac{1}{2\lambda^2} + \frac{2NK_s \sin^2 f}{r^2}\right)\left(\frac{\partial f}{\partial r}\right)^2 + \frac{B^2 N \sin^2 f}{2\lambda^2 r^2} \tag{18}$$

$$T_{\theta\theta} = -\frac{C^2}{2B^2}\left(\frac{1}{\lambda^2} + \frac{4NK_s \sin^2 f}{r^2}\right)\left(\frac{\partial f}{\partial r}\right)^2 + \left(n^2 + \frac{C^2 N}{2r^2}\right)\frac{\sin^2 f}{\lambda^2} \tag{19}$$

$$T_{\theta z} = \frac{\omega}{2B^2}\left(\frac{1}{\lambda^2} + \frac{4NK_s \sin^2 f}{r^2}\right)\left(\frac{\partial f}{\partial r}\right)^2 + \left(nmk - \frac{\omega N}{2r^2}\right)\frac{\sin^2 f}{\lambda^2} \tag{20}$$

$$T_{zz} = -\frac{1}{2B^2 C^2}\left(1 + \frac{\omega^2}{r^2}\right)\left(\frac{r^2}{\lambda^2} + 4NK_s \sin^2 f\right)\left(\frac{\partial f}{\partial r}\right)^2 + \left(m^2 k^2 + \frac{N}{2C^2}\left(1 + \frac{\omega^2}{r^2}\right)\right)\frac{\sin^2 f}{\lambda^2} \tag{21}$$

where $N = n^2 g_{zz} - 2nmk g_{\theta z} + m^2 k^2 g_{\theta\theta}$. Note that $T_{\theta z}$ is in general non-zero, provided that either $mk$ or $\omega$ is non-zero. In fact, $mk$ acts as a source term for $\omega$. The twist in the vortex is therefore solely responsible for a non-zero circular stress, $T_{\theta z}$.

In components form, the Einstein field equations (11), corresponding to the metric tensor are

$$G_{tt} = R_{tt} - \frac{A^2 R}{2}, \; G_{rr} = R_{rr} + \frac{B^2 R}{2}, \; G_{\theta\theta} = R_{\theta\theta} + \frac{C^2 R}{2}, \; G_{\theta z} = R_{\theta z} - \frac{\omega R}{2}, \; G_{zz} = R_{zz} + \frac{(r^2 + \omega^2)}{2C^2} R \tag{22}$$

## 3. Solution of the Einstein Field Equations

We use ode45 for numerical calculation, i.e. a tool box in Matlab to solve ordinary differential equations (ODE). Such tool box works based on explicit formula of 4th, 5th order Runge-Kutta. Solution scheme with ode45 (or in general, Runge-Kutta) is transforming one or more than one differential equations of $n$ order into $n$ differential equations of first order. In order to obtain numerical solution, we need $n$ suitable initial values for $n$ differential equations of first order. Perhaps, this is the reason why this method is called an initial value method, and system of equations which are solved called an initial value problem (IVP) [4].

The equations (22) can be rearranged as source equations for the first derivatives of $B$ and $f$ and the second derivatives of $A$, $C$, $\omega$ as follows [4].

$$B' = \frac{BA'}{A} + \frac{r\varepsilon B^3}{\lambda^2}\left[\frac{n^2}{C^2}\left(1 + \frac{\omega^2}{r^2}\right) + \frac{2nmk\omega}{r^2} + \frac{C^2 m^2 k^2}{r^2}\right]\sin^2 f \tag{23}$$

$$f' = \pm\left(\frac{\varepsilon}{2\lambda^2} - 2\varepsilon K_s\left[\frac{n^2}{C^2}\left(1 + \frac{\omega^2}{r^2}\right) + \frac{2nmk\omega}{r^2} + \frac{C^2 m^2 k^2}{r^2}\right]\sin^2 f\right)^{-1/2}$$

$$\times \left(\begin{array}{l} \dfrac{A'}{rA} + \dfrac{\omega\omega' C'}{r^2 C} + \dfrac{C'}{rC} - \dfrac{C'^2}{C^2}\left(1 + \dfrac{\omega^2}{r^2}\right) - \dfrac{\omega'^2}{4r^2} \\ + \dfrac{\varepsilon B^2}{2\lambda^2}\left[\dfrac{n^2}{C^2}\left(1 + \dfrac{\omega^2}{r^2}\right) + \dfrac{2nmk\omega}{r^2} + \dfrac{C^2 m^2 k^2}{r^2}\right]\sin^2 f \end{array}\right)^{1/2} \tag{24}$$

$$A'' = -\frac{A'}{r} + \frac{A'^2}{A} + \varepsilon \sin^2 f \left[ \frac{n^2}{C^2}\left(1+\frac{\omega^2}{r^2}\right) + \frac{2nmk\omega}{r^2} + \frac{C^2 m^2 k^2}{r^2} \right]$$

$$\times \left\{ \frac{rA'B^2}{\lambda^2} - 2K_s \left( \begin{array}{l} \dfrac{A'}{r} + \dfrac{\omega A \omega' C'}{r^2 C} + \dfrac{AC'}{rC} - \dfrac{AC'^2}{C^2}\left(1+\dfrac{\omega^2}{r^2}\right) - \dfrac{A\omega'^2}{4r^2} \\ + \dfrac{\varepsilon AB^2}{2\lambda^2}\left[\dfrac{n^2}{C^2}\left(1+\dfrac{\omega^2}{r^2}\right) + \dfrac{2nmk\omega}{r^2} + \dfrac{C^2 m^2 k^2}{r^2}\right]\sin^2 f \end{array} \right) \right.$$

$$\left. \times \left( \frac{\varepsilon}{2\lambda^2} - 2\varepsilon K_s \left[\frac{n^2}{C^2}\left(1+\frac{\omega^2}{r^2}\right) + \frac{2nmk\omega}{r^2} + \frac{C^2 m^2 k^2}{r^2}\right] \sin^2 f \right)^{-1} \right\} \quad (25)$$

$$C'' = \frac{C'^2}{C}\left(1 + \frac{2\omega^2}{r^2}\right) - \frac{2\omega\omega' C'}{r^2} - \frac{C'}{r} + \frac{C\omega'^2}{2r^2}$$

$$- \frac{\varepsilon B^2 \sin^2 f}{\lambda^2}\left[\frac{n^2}{C^2}\left(1+\frac{\omega^2}{r^2}\right) + \frac{2nmk\omega}{r^2} + \frac{C^2 m^2 k^2}{r^2}\right]\left(\frac{C}{2} - rC'\right)$$

$$+ \frac{\varepsilon \sin^2 f}{2}\left[\frac{n^2}{C^2}\left(1+\frac{\omega^2}{r^2}\right) + \frac{2nmk\omega}{r^2} + \frac{C^2 m^2 k^2}{r^2}\right]$$

$$\times \left\{ \frac{CB^2}{\lambda^2} - 4K_s \left( \begin{array}{l} \dfrac{A'C}{rA} + \dfrac{\omega\omega' C'}{r^2} + \dfrac{C'}{r} - \dfrac{C'^2}{C}\left(1+\dfrac{\omega^2}{r^2}\right) - \dfrac{\omega'^2 C}{4r^2} \\ + \dfrac{\varepsilon B^2 C \sin^2 f}{2\lambda^2}\left[\dfrac{n^2}{C^2}\left(1+\dfrac{\omega^2}{r^2}\right) + \dfrac{2nmk\omega}{r^2} + \dfrac{C^2 m^2 k^2}{r^2}\right] \end{array} \right) \right.$$

$$\left. \times \left( \frac{\varepsilon}{2\lambda^2} - 2\varepsilon K_s \sin^2 f \left[\frac{n^2}{C^2}\left(1+\frac{\omega^2}{r^2}\right) + \frac{2nmk\omega}{r^2} + \frac{C^2 m^2 k^2}{r^2}\right] \right)^{-1} \right\} \quad (26)$$

$$\omega'' = \frac{\omega'}{r} + \frac{4\omega C'^2}{C^2}\left(1+\frac{\omega^2}{r^2}\right) - \frac{4\omega^2 \omega' C'}{r^2 C} - \frac{4\omega C'}{rC} + \frac{\omega\omega'^2}{r^2}$$

$$- \frac{\varepsilon B^2 (\omega - r\omega')\sin^2 f}{\lambda^2}\left[\frac{n^2}{C^2}\left(1+\frac{\omega^2}{r^2}\right) + \frac{2nmk\omega}{r^2} + \frac{C^2 m^2 k^2}{r^2}\right]$$

$$+ \varepsilon \sin^2 f \left[\frac{n^2}{C^2}\left(1+\frac{\omega^2}{r^2}\right) + \frac{2nmk\omega}{r^2} + \frac{C^2 m^2 k^2}{r^2}\right]$$

$$\times \left\{ \frac{\omega B^2}{\lambda^2} - 4K_s \left( \begin{array}{l} \dfrac{\omega A'}{rA} + \dfrac{\omega^2 \omega' C'}{r^2 C} + \dfrac{\omega C'}{rC} - \dfrac{\omega C'^2}{C^2}\left(1+\dfrac{\omega^2}{r^2}\right) - \dfrac{\omega\omega'^2}{4r^2} \\ + \dfrac{\varepsilon \omega B^2 \sin^2 f}{2\lambda^2}\left[\dfrac{n^2}{C^2}\left(1+\dfrac{\omega^2}{r^2}\right) + \dfrac{2nmk\omega}{r^2} + \dfrac{C^2 m^2 k^2}{r^2}\right] \end{array} \right) \right.$$

$$\left. \times \left( \frac{\varepsilon}{2\lambda^2} - 2\varepsilon K_s \sin^2 f \left[\frac{n^2}{C^2}\left(1+\frac{\omega^2}{r^2}\right) + \frac{2nmk\omega}{r^2} + \frac{C^2 m^2 k^2}{r^2}\right] \right)^{-1} \right\} \quad (27)$$

The sign ($\pm$) at (24) suggest us to choose it so that the values $f' < 0$ for most or all values of $r$. Here we choose the sign (-). Equation (23) to (27) need to be solved simultaneously. To do so, we convert second order ODE from (25) to (27) to be first order ODE. At this stage one finally get eight coupled first order ODEs which are ready to solve.

We obtain equations to generate initial value for $r = r_{min}$, i.e. [4]

$$B(r_{min}) = B_0 + \frac{\varepsilon a^2 B_0}{2}\left(\frac{1}{\lambda^2} - \frac{K_s a^2}{B_0^2}\right) r_{min}^2 \tag{28}$$

$$f(r_{min}) = \pi + a r_{min} \tag{29}$$

$$A(r_{min}) = A_0 - \frac{\varepsilon K_s a^4 A_0}{2 C_0^2} r_{min}^2 \tag{30}$$

$$C(r_{min}) = B_0 r_{min} + \frac{\varepsilon K_s a^4}{2 B_0} r_{min}^3 \tag{31}$$

$$\omega(r_{min}) = \omega_1 r_{min}^2 \tag{32}$$

where $B_0$, $a$, $A_0$, $\omega_1$, $K_s$ and $\lambda$ are unknown constants. The values assumed for $B_0$ and $A_0$ affect the solution only by rescaling the metric functions, the radial coordinate $r$ and the twist term $mk$. This follows from the fact that:

i.  All the components of the Einstein tensor except $G_{tt}$ in (22) depend on $A$ only through $A''/A$ or $A'/A$, while $G_{tt}$ is proportional to $A^2$, and similarly all the components of the stress-energy tensor except $T_{tt}$ in (17) to (21) do not depend on $A$, while $T_{tt}$ is proportional to $A^2$, so we can multiply $A$ by any non-zero constant without affecting any of the other metric functions or $f$.

ii. A transformation of $r \to cr$, $\frac{d}{dr} \to \left(\frac{1}{c}\right)\frac{d}{dr}$, $\frac{d^2}{dr^2} \to \left(\frac{1}{c^2}\right)\frac{d^2}{dr^2}$, $B \to \left(\frac{1}{c}\right)B$, $\omega \to c\omega$, $mk \to cmk$, where $c$ is any positive constant, leaves the field equations (23) to (27) unchanged.

So, it is always possible to divide $A$ by $A_0$ and $B$ by $B_0$ so that the initial values of $A$ and $B$ (at $r = 0$) are both 1. And in doing this, $C_0$ is also rescale to 1, as $C = C_0 r + ...$ is unchanged, and so if $r \to B_0 r$ we must have $C_0 \to C_0/B_0 = 1$. Making this rescaling simplifies equations (28) – (32), i.e

$$B(r_{min}) = 1 + \frac{\varepsilon a^2}{2}\left(\frac{1}{\lambda^2} - K_s a^2\right) r_{min}^2 \tag{33}$$

$$f(r_{min}) = \pi + a r_{min} \tag{34}$$

$$A(r_{min}) = 1 - \frac{\varepsilon K_s a^4}{2} r_{min}^2 \tag{35}$$

$$C(r_{min}) = r_{min} + \frac{\varepsilon K_s a^4}{2} r_{min}^3 \tag{36}$$

$$\omega(r_{min}) = \omega_1 r_{min}^2 \tag{37}$$

For $r \to \infty$, no matter of our choices for the values of such unknown constants, we want to graph $B(r)$, $f(r)$, $A(r)$, $C(r)/r$ and $(r^2 + \omega^2)/C^2$ solutions respectively as graphics which are asymptotic to $B \to 1$, $f \to 0$, $A \to 1$, $C/r \to (const. \neq 0)$, and $(r^2 + \omega^2)/C^2 \to 1$. If we define $r_{min} = 0.0001$, $r_{max} = 4 \times 10^4$, $\varepsilon = 8\pi$, $a = -0.186$, $\omega_1 = 10^{-15}$, $K_s = 4 \times 10^{-14}$, $\lambda = 2100$, $n = 1$, $mk = 10^{-6}$, we obtained graphs of typical solution as follows

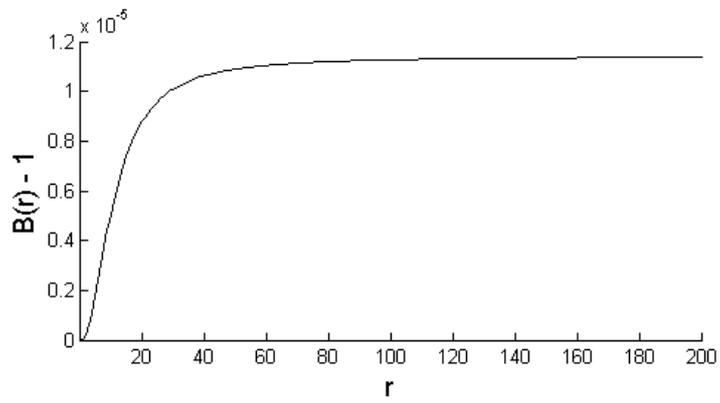

**Figure 1** Typical form of the $B(r)-1$ function

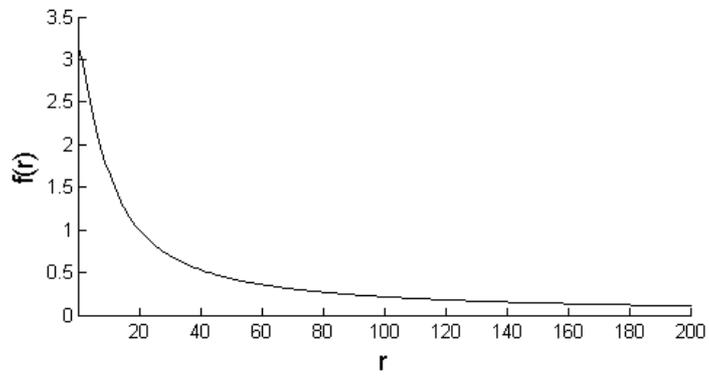

**Figure 2** Typical form of the $f(r)$ function

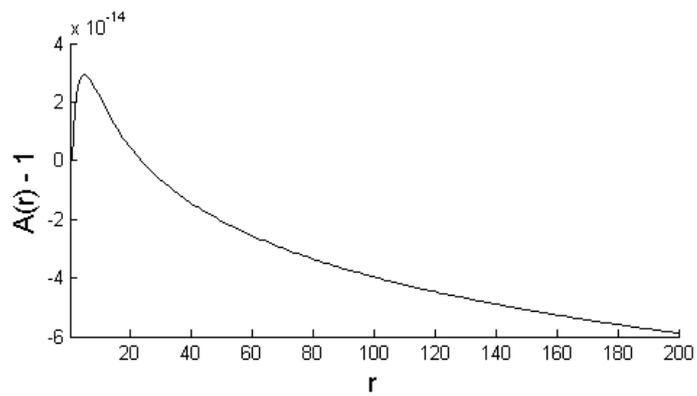

**Figure 3** Typical form of the $A(r)-1$ function

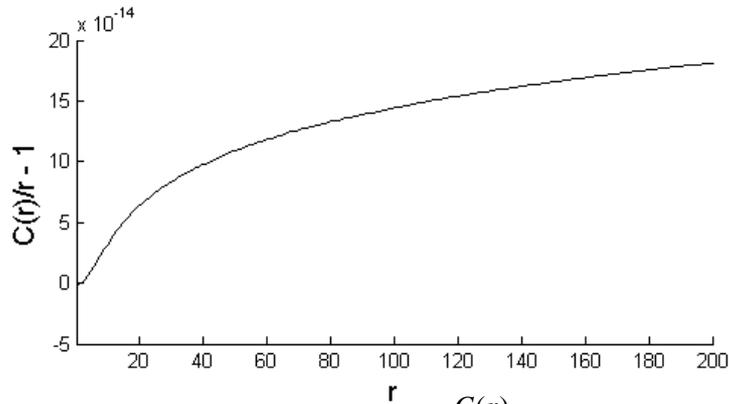

**Figure 4** Typical form of the $\dfrac{C(r)}{r} - 1$ function

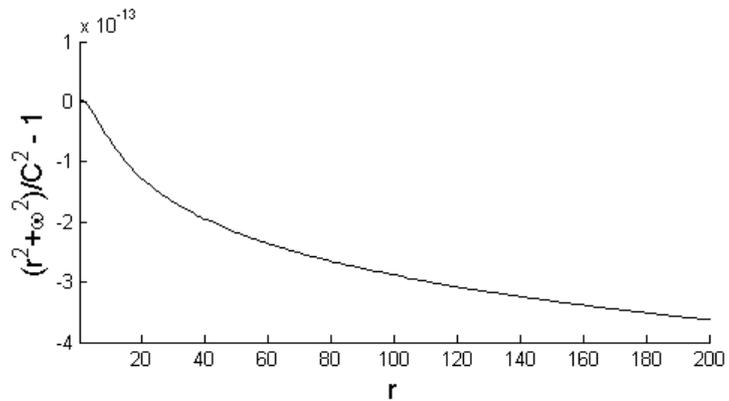

**Figure 5** Typical form of the $\dfrac{r^2 + \omega^2}{C^2} - 1$ function

**4. Acknowledgment**
This research is supported fully by Graduate Research Scholarship Universiti Brunei Darussalam (GRS UBD). This support is highly appreciated.